\documentclass[11pt]{article}
\evensidemargin=\oddsidemargin

\usepackage{graphicx}
\usepackage{epsfig}    
\usepackage{dcolumn}
\usepackage{bm}
\usepackage{amssymb}
\usepackage{epsfig}    
\usepackage{here}

\renewcommand{\thefootnote}{\fnsymbol{footnote}}

\def\beqn{\begin{eqnarray}}
\def\eeqn{\end{eqnarray}}
\relax
\newcommand{\ba}[1]{\begin{array}{#1}}
\def\ea{\end{array}}

\def\beq{\begin{equation}}
\def\eeq{\end{equation}}
\def\bea{\begin{array}}
\def\eea{\end{array}}

\def\to{\rightarrow}

\def\dis{\displaystyle}
\def\f{\frac}
\def\[{\left[}
\def\]{\right]}
\def\({\left(}
\def\){\right)}

\def\la{{\lambda}}
\def\ep{{\epsilon}}
\def\qs{{\sqrt{2}}}

\def\sm0{{\widetilde{m}_0}}

\def\tas{\tilde{\tau}}
\def\mus{\tilde{\mu}}
\def\slp{\tilde{l}}

\def\cut{{\Lambda}}

\def\U1em{{U(1)_{\rm em}}}
\def\to{\rightarrow}

\def\sq2{\sqrt{2}}

\def\End{\end{document}}
\newcommand{\lae}{\stackrel{<}{\sim}}

\begin{document}                                                              
\thispagestyle{empty}


\begin{center}
{\bf {\large
Flavor issues in the Higgs Sector}}

\vspace*{15mm}

{\sc\large J. Lorenzo Diaz-Cruz}
\vspace*{4mm}  \\
~Instituto de Fisica, BUAP\\
Ap. Postal J-48, Puebla, Pue., 72570\\
 Mexico
\end{center}

\vspace*{15mm}
\begin{abstract}
\hspace*{-0.35cm}
We discuss the conditions under which the flavor structure 
of SUSY model induces, either radiatively or through 
mixing, new flavor-violating interactions in the Higgs sector. 
The {\it{radiative}} flavor mediation mechanism is illustrated 
using the minimal SUSY extension of the SM (MSSM) with generic 
trilinear A-terms, and applied to evaluate the  corrections 
to Lepton Flavor-Violating (LFV) and Flavor-Conserving (LFC)
Higgs vertices. Flavor mediation through {\it{mixing}} is discussed 
within the context of an $E_6$-inspired multi-Higgs model, 
suplemented with an abelian flavor symmetry. Tevatron 
and LHC can probe the flavor structure of these models through the 
detection of the LFV Higgs mode $h\to \tau \mu $, while
NLC can perform high-precision tests
of the LFC mode $h \to \tau^+ \tau^-$.


\end{abstract}


\setcounter{page}{1}
\setcounter{footnote}{0}
\renewcommand{\thefootnote}{\arabic{footnote}}

\noindent
{\bf {\large 1. Introduction.}}
The discovery of atmospheric neutrino oscillations \cite{superkam},
as well as the new meassurements of CP-violation \cite{nirtalk}, 
can be considered some of the most important recent results in 
particle physics, and  
toghether with the reported bounds on the Higgs boson mass
\cite{hsmradc}, are helping us to shape our understanding of 
flavor physics and electroweak symmetry breaking. 
The Higgs boson mass is constrained by radiative corrections to lay 
in the range 110-185 GeV at 95 \% c.l. \cite{hsmradc}; such
a light higgs boson is consistent with the 
predictions from  weak-scale supersymmetry (SUSY), 
which has become one of the leading candidates for physics
beyond the standard model.  The Higgs sector of the
MSSM includes two higgs doublets, and the light Higgs boson
(with mass bound $m_h \lae 125$ GeV), is perhaps 
the strongest prediction of the model.

 However, after a Higgs signal will  be seen, probably
at the Tevatron and/or LHC, it will become crucial 
to meassure its mass, spin and couplings, 
to elucidate its nature. In particular, the Higgs coupling 
to light fermions ($b\bar{b}, c\bar{c}, \tau^+\tau^-$)
could be measured at next-linear collider (NLC) with a precision 
of a few percent, which can be used to constrain physics beyond 
the SM.  For instance, higher-dimensional operators of the type  
$\Phi^\dagger \Phi \bar{Q}_L \Phi b_R$
involving the third family,  will generate corrections
to the coupling $h\bar{b}b$, which in turn will
modify the dominant decay of the light Higgs,
as well as the associated production of the Higgs with 
b-quark pairs \cite{ourhixbb, carena}.

 The most widely studied scenarios for Higgs searches, assume that
the Flavor-Conserving (FC) Higgs-fermion couplings only depend on 
the diagonalized fermion mass matrices, while flavor-violating (FV) 
Higgs transitions are absent or highly suppressed. Indeed, within the SM 
the Higgs boson-fermion couplings are only sensitive 
to the fermion mass eigenvalues. However, if one considers extensions 
of the SM, which either present a significant source of flavor-changing 
transition  or are aimed precisely to explain the pattern of masses and 
mixing angles of the quarks and leptons, then it is quite possible that
such physics will include new flavored interactions.  Namely, when 
additional fields that have non-aligned couplings to the SM fermions, 
i.e. which are not diagonalized by the same rotations that
diagonalize the fermion mass matrices, also couple
to the Higgs boson, then such fields could be responsible for
transmiting the structure of the flavor sector to the  Higgs 
bosons, thereby producing a {\it{a more flavored Higgs boson}}. 

As a consequence of the presence of  Lepton Flavor Violating (LFV) 
higgs interactions, the decay $h \to \tau\mu$ can be induced at rates 
that could be detected at future colliders, which could be the 
manifestation of a deeper link between the Higgs and flavor sectors.
In fact, a large coupling $h\tau\mu$ is suggested by the large 
$\nu_\mu-\nu_\tau$ mixing observed with atmospheric neutrinos.
The importance of LFV Higgs modes has been discussed in refs.
\cite{ourhlfv,prihmutau},  and it is also the main
focus of this work.
Depending on the nature of such new physics, we can 
identify two possibilities for flavor-Higgs mediation, namely:
\begin{enumerate}  
\item {\it{RADIATIVE MEDIATION.}} In this case the Higgs sector has 
diagonal couplings to the fermions at tree-level.
However, in the presence of new particles associated with extended 
flavor physics, which couple both to the Higgs and to the SM fermions,
these flavor-mediating fields will induce corrections to the Yukawa
couplings and/or new FCNC proccess at loop levels.
This case will be  discussed within the context of the MSSM
with general trilinear soft-breaking terms. 
\item {\it{MIXING MEDIATION.}} Modifications to the Higgs-flavor
structure can also arise when additional particles (bosons or fermions)
mix with the SM ones. These new interactions could then be transmited 
to the Higgs sector, either through scalar-Higgs mixing or through
mixing of SM fermions with exotic ones. A multi-Higgs $E_6$-inspired 
model, supplemented  with an abelian-flavor symmetry, 
will be used as an example of this mechanism.
\end{enumerate}
Other models where flavor-higgs mediation could occur
through mixing include:
i) the general two-Higgs doublet model (THDM-III) 
\cite{hifcnc,mythdiiia} ,
ii) A model where the SM fermions mix with mirror fermions
\cite{ourhlfv} , as well as the models with:
iii) Higgs-flavon mixing, and iv) R-parity breaking scenarios.

{\bf {\large 2. Flavor-mediation in the MSSM.}}
Within the MSSM, it can be shown that flavor-Higgs mediation
is of radiative type, and it communicates the non-trivial flavor
structure of the soft-breaking sector  to the  Higgs bosons
through gaugino-sfermion loops. 
As an illustration of this case, we have evaluated the 
SUSY contributions to the Higgs-lepton vertices,
arising from the slepton mixing that originates from the
trilinear $A_l$-terms. The slepton mixing is 
constrained by the low-energy data, but it 
mainly suppress the FV's associated with the first two 
family sleptons, and still allows the flavor-mixings 
between the second- and third-family sleptons, 
to be as large as $O(1)$. Thus, one can neglect the mixing 
involving the selectrons, and the general $6\times 6$ 
slepton-mass-matrix reduces down to a $4\times 4$ matrix,
involving only the smuon ($\mus$) and stau ($\tas$) sectors, 
similarly to the squarks case discussed in ref. \cite{oursqmix}.  
Such pattern of large slepton mixing, can also be motivated
by considering GUT models that incorporate the large neutrino 
mixing observed with atmospheric neutrinos.

To evaluate the loop corrections both to the 
the lepton-flavor conserving (LFC) ($h \to l_i l_i$) and
the flavor violating (LFV) Higgs modes ($h\to l_i l_j $),
we have performed the diagonalization of the slepton 
mass matrices, and expressed the gaugino-lepton-slepton
and higgs-slepton-slepton interactions in the mass
eigenstate basis, thus without relying on the mass-insertion 
approximation. The results for the radiative
corrections to the LFC modes are in general small,
and can be neglected here \cite{myMFHB}. While
the decay width for $h\to l_i l_j $
(adding both final states $l^+_i l^-_j$ and $l^-_i l^+_j$ )
is written as:

\newpage 

\begin{equation}
\Gamma (h \to {l_i l_j}) \, = \,
{ \frac{{m_h}}{ 8 \pi } } ( |F_L|^2+|F_R|^2 ) 
\end{equation}

Including only the vertex corrections, $F_{L,R}$ 
are given by:
\beqn
\dis 
F^{V}_{L} & \!\!=\!\! & \frac{g^2_1 m_{\tilde{B}} }{32 \pi^2}                   
         \!\sum_{\alpha\beta}\!
            \lambda^L_{jk} C_0
           (m_h^2,m_{\tau}^2,0; m_{\tilde{l}_{\alpha}},
            m_{\tilde{B}}, m_{\tilde{l}_{\beta}})\,,
\nonumber
\\[-1mm]
\label{eq:FVERch}
\\[-2mm]
\dis
F^V_{R} & \!\!=\!\! & \frac{g^2_1 m_{\tilde{B}} }{32 \pi^2}                  
         \!\sum_{\alpha \beta}\!
            \lambda^R_{\alpha \beta} C_0
           (m_h^2,m_t^2,0; m_{\tilde{l}_{\alpha}},
            m_{\tilde{B}}, m_{\tilde{l}_{\beta}})\,,
\nonumber
\eeqn
where 
$\tilde{l}_{\alpha,\beta}\in (\mus_1,\mus_2,\tas_1.\tas_2)$,
$C_0$ denotes the 3-point $C$-function of Passarino-Veltman.
$\lambda^{L,R}_{\alpha k}$ is the product of the relevant
$h\slp_{\alpha}\slp_{\beta}$ and
$\slp_{\alpha}\tilde{B}\tau(\mu)$ couplings,
(For details see Ref. \cite{myMFHB}).
As shown in table 1, the LFV mode $h\to \tau \mu$ has a branching ratio
that may reach the $4\times 10^{-4}$; this result may be at the 
reach of LHC (as one can conclude by comparing with
the minimum detectable B.r. \cite{ourhlfv,prihmutau}).

\bigskip

\bigskip

\noindent
{Table\,1. 
Br$[ h \to \tau \mu]$
is shown for a sample set of SUSY inputs with 
$(\mu,m_A)=(0.2,0.3)$\,TeV, $A=\f{\sm0}{2}$
and $\tan\beta=5 (10)$.
The numbers in each entry are obtained using the maximum
value $x_{max} (\simeq 1.2-3.0)$ allowed for the
given set of SUSY parameters.}
\vspace*{1.5mm}
\begin{center}
\begin{tabular}{|c|c|c|}
\hline
$m_{\tilde B}$ & $\sm0=450$ GeV & $\sm0=600$ GeV \\ 
\hline
 150 GeV   & $1.1 \times 10^{-7}$ ($3.0 \times 10^{-8}$)
           & $5.0 \times 10^{-5}$ ($1.2 \times 10^{-5}$)  \\
\hline
 300 GeV   & $3.1 \times 10^{-7}$ ($8.0 \times 10^{-8}$)
           & $8.0 \times 10^{-5}$ ($2.1 \times 10^{-5}$)   \\
\hline
 600 GeV   & $5.3 \times 10^{-5}$ ($1.4 \times 10^{-5}$)
           & $4.4 \times 10^{-4}$ ($1.2 \times 10^{-4}$)   \\
\hline
\end{tabular}
\end{center}

\bigskip

{\bf {\large 3. An $E_6$-inspired multi-Higgs model.}}
On the other hand, flavor-Higgs mediation through 
mixing occurs within an $E_6$-inspired  multi-Higgs model \cite{sheretal},
suplemented with an abelian flavor symmetry.
 Large Higgs-FV  effects are also found to arise, though in this case 
at the tree-level; in this model there is a Higgs pair associated with 
each family. Then, to generate a realistic 
flavor structure for both leptons and sleptons
we include a horizontal $U(1)_H$ symmetry, which
at the same time helps to keep under control the FCNC problem, 
via proper powers of a single suppression factor
$\ep=\langle S\rangle/\cut$ \,\cite{U1H},
which has a similar size as the Wolfenstein-parameter $\la$ in
the CKM matrix, 
i.e., $\ep\simeq \la \simeq 0.22$\,\cite{U1H}.
Here, $\langle S\rangle$ denotes the vacuum expectation
value of a singlet scalar $S$, responsible for 
spontaneous $U(1)_H$ breaking,
and $\Lambda$ is the scale at which the $U(1)_H$ breaking is 
mediated to light fermions.
The Yukawa lagrangian is written as:

\begin{equation}
{\cal{L}}_Y= \bar{U}_i Y^u_{ij} H^u_\alpha Q_j -
             \bar{D}_i Y^d_{ij} H^d_\alpha Q_j -
              \bar{E}_i Y^l_{ij} H^d_\alpha L_j 
\end{equation}
where $H^{u,d}_\alpha$ ($\alpha=1,2,3$) denote the three Higgs pairs
of the model. Then, assuming that all Higgs
pairs have vanishing charges under the flavor symmetry
$U(1)_F$, we can induce Yukawa couplings that satisfy current
data on quark and lepton masses, as well as CKM angles.
Working in the basis where only $H^{u,d}_3= H_{u,d}$ 
aquires a v.e.v. ($<H^0_{u,d}>=v_{u,d}$), 
and considering only the two-flavor tau-mu case,  
we can write the charged lepton mass matrix 
($M_l= \f{v_d}{\sqrt{2}} Y^l$), 
by assigning the flavor-charges: 
$(h_2,h_3)= (2,2)$ and $(\beta_2,\beta_3)= (3,1)$,
to the lepton doublet and singlet, respectively,
then:
\beq
M_l \,\sim\, \dis \f{v_d}{\qs}
\left\lgroup
            \bea{cc}
             \la^5  &  \la^5 \\
              \la^3     &  \la^3  
            \eea
\right\rgroup ,
\label{eq:Mlept}
\eeq
which gives the correct order of magnitude for the
charged lepton masses, nameley 
$m_\mu \simeq m_\tau \la^2 \simeq \la^5 v_d$.
Then, the ``Yukawa matrices'' that describe the 
interactions for the remmaining  Higgs doublets $H^d_{1,2}$,
induce LFV and LFC interactions through the mixing of heavy states
with the light MSSM-like Higgs, and can be described as
follows:
\begin{equation}
{\cal{L}}_{int}=  
\f{g m_\tau \sin\alpha}{\sqrt{2}m_W \cos\beta}
[-\epsilon_l\f{(1-z_1)}{\sin\alpha} \bar{\tau} \mu
+ (1-\epsilon_l\f{(1+z_1)}{\sin\alpha}) \bar{\tau}  \tau +h.c.] h^0
\end{equation}
where  $z_{1,2}$ correspond to the 
$O(1)$ coefficients left undetermined by the FN approach,
and $\epsilon_l$ is used to parametrize the mixing between 
the light MSSM-like Higgs and the heavier Higgs states. 

Then, to evaluate the Higgs-FV interactions for tau-mu and the
corrections to the FC Higgs-tau vertex, we have considered 
the values $z_1=0.75,0.9$ and $\epsilon_l=0.1, .05$. 
From table 2, we can see that the decay branching ratio 
${\rm Br}[h\to \tau \mu]$ can be of the order $10^{-2}-10^{-3}$, 
over the part of the SUSY parameter space with large 
values of $\tan\beta$,  and when the mass of the lightest 
Higgs boson $h^0$ is around $115-120$ \,GeV, which can be 
detected at future colliders. In fact, the rates obtained in this model 
for $z_1=0.75$ and $\tan\beta \geq 20$ are at the reach of Tevatron 
Run-2, while LHC can have a larger sensitivity to discover this LFV
mode $h\to \tau\mu$ in largest portions of parameter
space.

  On the other hand, this model also predicts corrections to
the Higgs-tau couplings, which can be tested at NLC. Table 2 
shows the resulting deviation of the Higgs width ($h\to \tau^+\tau^-$) 
from the MSSM value, defined as:
$\Delta \Gamma_{h\tau\tau}= \f{\Gamma_{h_{E_6}\tau\tau} }
                             {\Gamma_{h_{MSSM}\tau\tau} }$
This table shows that $\Delta \Gamma_{h\tau\tau}$ can easily be above 0.08, 
which according to current studies, could be measurable at the
NLC.

\bigskip

\bigskip

\noindent
{Table\,3.  Values of $B.R.(h\to \tau\mu)$ and 
$\Delta \Gamma_{h\tau\tau}$ that arise for
$z_1=0.75,0.9$ and $\epsilon_l=0.1$.
Results in each parenthesis correspond to $\tan\beta= 5,10,20$}
\vspace*{1.5mm}
\begin{center}
\begin{tabular}{|c|c|c|c|}
\hline
 $m_A$    &  $z_1$    & $B.R.(h\to \tau\mu)\times 10^3$  
&  $\Delta \Gamma_{h\tau\tau}$ \\
\hline
 100 GeV  &  0.75      &  (0.19, 0.16, 0.15)
                       &  (0.69, 0.72, 0.74)        \\
          &  0.90      &  (0.03, 0.027, 0.024)
                       &  (0.66, 0.69, 0.71)        \\
\hline
 150 GeV  &  0.75      &  (0.64, 0.17,  0.56)
                       &  (0.44, 0.29,  0.04)        \\
          &  0.90      &  (0.10, 0.27,  0.90)
                       &  (0.40, 0.15,  0.01)        \\
\hline
 200 GeV  &  0.75      &  (1.40, 4.80,  17.0)
                       &  (0.23, 0.03,  0.95)        \\
          &  0.90      &  (0.22, 0.76,  2.70)
                       &  (0.19, 0.07,  1.30)        \\
\hline
 250 GeV  &  0.75      &  (1.90, 7.20,  15.0)
                       &  (0.13, 0.06,   2.0)        \\
          &  0.90      &  (0.31, 1.10,  3.90)
                       &  (0.10, 0.13,  2.60)        \\
\hline
 300 GeV  &  0.75      &  (2.40, 8.80,  29.0)
                       &  (0.09, 0.16,  2.80)        \\
          &  0.90      &  (0.38, 1.40,  4.60)
                       &  (0.05, 0.27,  3.50)        \\
\hline
\end{tabular}
\end{center}

\bigskip

\bigskip

{\bf {\large 4. Conclusions.}}
We have discussed the conditions under which the flavor structure 
of SUSY model induces, either radiatively or through 
mixing, new flavor-violating interactions in the Higgs sector, 
which can be probed at future high-energy colliders.
It is found that the  Higgs-FV  couplings
are induced at rates that can be significant enough to provide 
new discovery signals at the on-going Run-2 at the
Fermilab Tevatron Collider and the CERN Large Hadron Collider (LHC), 
which can detect the LFV mode $h\to \tau \mu $, 
and give information on the flavor structure of the model,
while NLC high-precision meassurements can bound the 
deviations from the SM for the LFC mode $h \to \tau^+ \tau^-$.
Implications for rare top quark decays \cite{mytcgama} are under
current investigation.



\end{document}